\documentclass[aip,jcp,reprint,floatfix]{revtex4-1}
\usepackage{graphicx}
\usepackage{amsmath}
\usepackage{color}
\begin{document}

\title{Dipole oscillator strength distributions, sum rules, mean excitation energies,
and isotropic van der {Waals} coefficients for benzene, pyridazine, pyrimidine,
pyrazine, $s$-triazine, toluene, hexafluorobenzene, and nitrobenzene}

\author{Ajit J. Thakkar}
\email{ajit@unb.ca}
\homepage{http://people.unb.ca/Ajit.Thakkar/}
\affiliation{Department of Chemistry, University of New Brunswick, Fredericton, New Brunswick, Canada E3B 5A3}%

\date{\today}

\begin{abstract}
Experimental, theoretical, and additive-model photoabsorption cross-sections combined
with constraints provided by the Kuhn-Reiche-Thomas sum rule and the high-energy
behavior of the dipole-oscillator-strength density are used to construct dipole
oscillator strength distributions for benzene, pyridazine (1,2-diazine), pyrimidine
(1,3-diazine), pyrazine (1,4-diazine), $s$-triazine (1,3,5-triazine), toluene
(methylbenzene), hexafluorobenzene, and nitrobenzene. The distributions are used to
predict dipole sum rules $S(k)$ for $-6 \le k \le 2$, mean excitation energies $I(k)$
for $-2 \le k \le 2$, and isotropic van der Waals $C_{6}$ coefficients.
A popular combination rule for estimating $C_{6}$ coefficients for unlike
interactions from the $C_{6}$ coefficients of the like interactions is found to be
accurate to better than 1\% for 606 of 628 cases (96.4\%) in the test set.
\end{abstract}

\maketitle

\section{Introduction}\label{sec:intro}

The current interest in non-covalent interactions has generated renewed efforts to
develop improved theoretical methods for the calculation of van der Waals
interactions.\cite{Stohr19} It is important to have reliable, experiment-based,
reference values of the van der Waals dispersion coefficients $C_{6}$ to assess these
theoretical methods. Since the early insight of Margenau,\cite{Margenau31} the most
reliable experiment-based values of $C_{6}$ coefficients have been obtained from
dipole oscillator strength distributions (DOSDs).

The DOSD of an atom or molecule\cite{Thakkar16DOSD} consists of the set of discrete
excitation energies $E_{i}$ and oscillator strengths $f_{i}$ together with the
differential dipole oscillator strength (DOS) function $(df/dE)$ for the continuum of
energies $E_{\text{c}}\le{E}<\infty$ that begins at the continuum threshold
$E_{\text{c}}$. The DOS is proportional to the photo-absorption cross section,
$\sigma$. Many constructions of a complete DOSD from experimental data possibly
subject to a few constraints have been reported; for a representative sample, see
Refs.~\onlinecite{Dalgarno67,Zeiss77a,Kumar96,Olney97,Berkowitz02}.

No complete DOSDs are available for (hetero)aromatic molecules apart from
benzene\cite{Kumar92a} and pyridine.\cite{Thakkar16DOSD} The purpose of this work is
to report the construction of constrained DOSDs and the resulting dipole sum rules
$S(k)$ for $-6 \le k \le 2$, mean excitation energies $I(k)$ for $-2 \le k \le 2$,
and isotropic van der Waals $C_{6}$ coefficients for pyridazine (1,2-diazine,
C$_{4}$H$_{4}$N$_{2}$), pyrimidine (1,3-diazine, C$_{4}$H$_{4}$N$_{2}$), pyrazine
(1,4-diazine, C$_{4}$H$_{4}$N$_{2}$), $s$-triazine (1,3,5-triazine,
C$_{3}$H$_{3}$N$_{3}$), toluene (methylbenzene, C$_{6}$H$_{5}$CH$_{3}$),
hexafluorobenzene (C$_{6}$F$_{6}$), and nitrobenzene (C$_{6}$H$_{5}$NO$_{2}$).
Moreover, an improved DOSD is reported for benzene (C$_{6}$H$_{6}$) because the
available DOSD\cite{Kumar92a} is now 28 years old and much new experimental
photoabsorption data has become available since then.

This work is organized as follows. Section~\ref{sec:methodology} is a summary of the
experimental data, additive models, and methods used to construct the DOSDs.
Section~\ref{sec:res} contains a discussion of the resulting DOSDs and molecular
properties. Section~\ref{sec:next} contains some thoughts about the future of DOSD
constructions.

\section{Methods and data}\label{sec:methodology}

\subsection{Experimental data}\label{sec:data}

Photoabsorption measurements for
benzene\cite{Rennie98,Etzkorn99,Feng02,Fally09,Capalbo16,Dawes17} more recent than
the 1992 DOSD construction\cite{Kumar92a} are available over the extended energy
range from 3.76$\,$eV to 200$\,$eV\@. Experimental photoabsorption data is available
for the other molecules only over a smaller energy range, roughly 4$\,$eV to
40$\,$eV\@. Fortunately, this energy range accounts for 90\% or more of the
polarizability.\cite{Jhanwar81,Kumar16HA} Experimental photoabsorption cross-sections
are available from 4.4$\,$eV to 40$\,$eV for pyridazine
(1,2-diazine),\cite{Holland13} from 3.6$\,$eV to 40$\,$eV for pyrimidine
(1,3-diazine),\cite{Bolovinos84,Silva10,Stener11} from 4.5$\,$eV to 40$\,$eV for
pyrazine (1,4-diazine),\cite{Bolovinos84,Stener11} from 3.9$\,$eV to 39$\,$eV for
$s$-triazine (1,3,5-triazine),\cite{Bolovinos84,Holland16} from 3.9$\,$eV to
35.6$\,$eV for
toluene,\cite{Bolovinos81,Bolovinos82,Hippler83,Shama91,Shaw98,Koban04,Eldib06,Serralheiro15}
and from 4.2$\,$eV to 41.3$\,$eV for
hexafluorobenzene.\cite{Philis81,Motch06,Holland09} For nitrobenzene, experimental
photoabsorption data is available\cite{Nagakura64,Shama91,Frosig00,Cooper01} for most
of the energy range from 3.3$\,$eV to 35.6$\,$eV but there is a gap from 8.1$\,$eV to
9.9$\,$eV\@. This gap was filled by forward extrapolation of Nagakura \textit{et
al.}'s data\cite{Nagakura64} and backward extrapolation of Cooper \textit{et al.}'s
data\cite{Cooper01} to their intersection point at roughly 8.4$\,$eV\@. Digital files
of most of the data mentioned in this paragraph were obtained either from the
MPI-Mainz UV/VIS Spectral Atlas\cite{KellerRudek13} or the Brion laboratory's
database.\cite{BrionDB}

Gas-phase measurements of the molar refractivity can be used as constraints for the
constructed DOSDs. However, they are available for only one of the eight molecules
considered in this work: benzene.\cite{Holemann1934,Ramaswamy1936}

\subsection{Additive models}\label{sec:additive}

Photoabsorption data is unavailable at higher energies: $E>200\,\text{eV}$ for
benzene, $E>35\,\text{eV}$ for toluene and nitrobenzene, and $E>40\,\text{eV}$ for
the five other molecules. Additive models are used to approximate this higher energy
data.

The simplest additive models are based on free atoms.\cite{*[{In the context of photoabsorption
cross-sections, additive models are sometimes called mixture rules. A succinct
history and description of additive models is given in }] [{. This paper set up a
hierarchy of additive models to help clarify what can be expected from a model in a
given level of the hierarchy.}] Thakkar12} For all the
molecules, photoabsorption cross-sections, $\sigma$, were constructed from each of
two free-atom additive models. One model, hereafter referred to as A(H), is based on
free-atom cross-sections taken from the semiempirical compilation of Henke \textit{et
al.}\cite{*[{}][{. The data is available digitally from
\url{https://henke.lbl.gov/optical_constants/asf.html} which also reports a few
updates to the published work.}] Henke93} that extends to 30,000$\,${eV}\@. The other
model, hereafter referred to as A(C), is based on the theoretical free-atom
cross-sections tabulated by Chantler\cite{Chantler95} for energies up to
100,000\,\text{eV}\@. Atom-additive models are expected to be sufficiently
accurate\cite{Henke93} for $E>70\,\text{eV}$ but have been used, in favorable
circumstances, for energies as low as 15$\,$eV\@.\cite{Jhanwar81}

More elaborate and presumably more accurate additive models\cite{Thakkar12} can be
constructed from molecular fragments (groups) provided that consistent data obtained
by the same experimental technique and preferably from the same laboratory is
available. Photoabsorption cross-sections, in the region from $25\,$eV to $200\,$eV,
were generated from two fragment-additive models for all the molecules except
benzene. The accuracy of a fragment-additive model depends partly on the choice of
the fragments. Chemical intuition can be used to help choose the fragments. For
example, the conceptual partition of the target species into fragments should
minimize disruption of rings and functional groups.

Since all the molecules studied in this work are closely related to benzene, it makes
sense to use benzene as one of the fragments and construct an appropriate correction
term from other fragments. Thus we express the cross-section $\sigma(\text{M})$ for
molecule M as
\begin{equation}\label{eq:fragment}
\sigma(\text{M})=\sigma(\text{benzene})+\Delta(\text{M})
\end{equation}
The azines are the most straightforward; bond additivity works reasonably well for
their polarizabilities.\cite{Doerksen96} Azines differ from benzene by $m=1,2,\dots$
aza-substitutions, that is substitutions of C--H by N. Fortunately, experimental
cross-sections in the desired energy range are available for both benzene and
pyridine, the simplest azine ($m=1$). The most chemically appealing fragment-additive
model used for the $m$-azines is:
\begin{equation}\label{eq:BP}
\Delta(m\text{-azine})=m[\sigma(\text{pyridine}) - \sigma(\text{benzene})]
\end{equation}
which is exact for benzene ($m=0$) and pyridine ($m=1$) and should be reasonably
accurate for the diazines ($m=2$) and triazines ($m=3$). Another fragment-additive
model considered for the azines is given by
\begin{equation}\label{eq:BNA}
\Delta(m\text{-azine})=\frac{m}{2}[\sigma(\text{N}_{2})
 -\sigma(\text{C}_{2}\text{H}_{2})] .
\end{equation}

The available data limits the accessible fragment models for the other three
molecules. One fragment-additive model for hexafluorobenzene is
\begin{equation}\label{eq:BMFM}
\Delta(\text{C}_{6}\text{F}_{6}) = 6[\sigma(\text{CH}_{3}\text{F}) - \sigma(\text{CH}_{4})]
\end{equation}
and another one is
\begin{equation}\label{eq:BTM}
\Delta(\text{C}_{6}\text{F}_{6}) =
\frac{3}{2}[\sigma(\text{CF}_{4})-\sigma(\text{CH}_{4})] .
\end{equation}
For toluene, the two fragment models considered are
\begin{equation}\label{eq:BAA}
\Delta(\text{toluene}) = \sigma(\text{propanone})-\sigma(\text{ethanal})
\end{equation}
and
\begin{equation}\label{eq:BEM}
\Delta(\text{toluene}) = \sigma(\text{C}_{2}\text{H}_{6})-\sigma(\text{CH}_{4}) .
\end{equation}
One fragment model used for nitrobenzene is
\begin{equation}\label{eq:BEMNO}
\Delta(\text{nitrobenzene}) = \frac{1}{2}\sigma(\text{C}_{2}\text{H}_{6})-\sigma(\text{CH}_{4})
 + \sigma(\text{NO}_{2}) .
\end{equation}
A fragment model based on a correction to pyridine rather than benzene is also used
for nitrobenzene:
\begin{equation}\label{eq:PCO2}
\sigma(\text{nitrobenzene}) = \sigma(\text{pyridine})+\sigma(\text{CO}_{2}) .
\end{equation}
For the sake of brevity, the fragment-additive model of Eq.~($m$) is referred to as
F($m$).

In this work, all the photoabsorption data used for fragment-additive models was
taken exclusively from the Brion database\cite{BrionDB} which contains data for 64
species. The data relevant to this work are unpublished data for propanone and
ethanal, and published data for benzene,\cite{Feng02} pyridine,\cite{Tixier02}
nitrogen,\cite{BrionN2} ethyne,\cite{BrionC2H2} methyl fluoride,\cite{BrionCH3F}
nitrogen dioxide,\cite{BrionNO2} tetrafluoromethane,\cite{BrionCF4} and methane and
ethane.\cite{*[{}][{. Erratum, \textit{ibid.}, \textbf{178}, 615 (1993).}]
BrionAlkane}

\subsection{DOSD construction}\label{sec:construction}

A robust method for constructing a DOSD from photoabsorption cross-sections combined
with constraints provided by the Kuhn-Reiche-Thomas (KRT) sum rule and usually molar
refractivity data was developed,\cite{Zeiss77a} refined, and applied to more than 50
species by Meath and coworkers as traced in Ref.~\onlinecite{Kumar08}. More recently,
the method was augmented by a high-energy constraint based on the asymptotic behavior
of the DOS. At first, this technique was restricted to homonuclear
molecules\cite{Kumar10,Kumar11c60} but it was later generalized to apply to all
molecules.\cite{Thakkar16DOSD} Since the initial application of the general method
to pyridine,\cite{Thakkar16DOSD} it has been applied to 20 more
molecules.\cite{Kumar16PCl,Kumar16HA,Kumar17}

A terse summary of this method suffices because a detailed description is available
elsewhere.\cite{Thakkar16DOSD} The available photoabsorption data is divided into
energy intervals $[E_{i},E_{i+1}]$ for $i=1,\dots,N$ in which $E_{1}$ is the
absorption threshold, $E_{N}$ is the highest energy for which a value of the DOS is
available, and $E_{N+1}=\infty$. Then a representative selection is made from the
initial distributions that can be constructed using different combinations of
experimental photoabsorption data from diverse sources, additive models, constraints,
and a three-term Laurent expansion for the asymptotic region $E>E_{N}$\@. The KRT sum
rule and the high-energy asymptotic behavior are always used as constraints. Initial
values of three parameters that appear in the high-energy
constraint\cite{Thakkar16DOSD} are obtained from Hartree-Fock values of the electron
density at the nucleus, $\rho(0)$, for each of the atoms in the molecule.\cite{*[{}]
[{.  $\rho(0)$ is calculated from the atomic number $Z$ and values of $b_{8}$ listed
in Table~2 of their work using the relation: $\rho(0)={\pi}b_{8}Z^{3}/8$.}] Koga96c}
Gas-phase molar refractivity values were used as low-energy constraints in some of
the distributions for benzene. For each selected distribution, iterations of a
constrained least-squares procedure are required to determine simultaneously two
parameters in the high-energy constraint\cite{Thakkar16DOSD} and the scale factors
$1+a_{i},\,i= 1,\dots,N$ for all the energy intervals. Finally, the best DOSD has to
be selected. Uniformity of the scale factors is a reflection of the consistency of
the initial data and so the distributions that lead to the smallest standard
deviations $s$ of the scaling parameters are considered the best. Almost invariably,
several DOSDs lead to values of $s$ very close to the lowest value, and then the
distribution leading to the smoothest DOS is selected from among these.

\subsection{Property calculations}\label{sec:props}

Once a final DOSD has been selected, quadrature is used to compute the dipole sums
$S(k)$ from
\begin{equation}\label{eq:sk}
S(k) = \int_{E_{\text{c}}}^{\infty} dE \left(\frac{df}{dE}\right)E^{k} ,
\end{equation}
the logarithmic sums $L(k)$ from
\begin{equation}\label{eq:lk}
L(k) = \int_{E_{\text{c}}}^{\infty} dE \left(\frac{df}{dE}\right) E^{k} \ln{E},
\end{equation}
and the mean excitation energies from:
\begin{equation}\label{eq:ik}
I(k)=\exp(\mathrm{d}\ln{S(k)}/\mathrm{d}k)=\exp(L(k)/S(k)) .
\end{equation}
Atomic units are used in the above and following equations. The expression used for
the isotropic dipole polarizability $\alpha(\omega)$ at selected frequencies $\omega$
is:
\begin{equation}\label{eq:alfa}
\alpha(\omega) = \int_{E_{\text{c}}}^{\infty} dE\, \frac{(df/dE)}{E^{2}-\omega^{2}}.
\end{equation}
Pseudospectral representations,\cite{Shohat1943}
$\{\epsilon_{i},f_{i},i=1,\dots,10\}$, of the DOSD are generated from the moments
$S(k)$, $-17\le {k}\le{+2}$. The pseudospectra are used to compute the $C_{6}$
coefficient for long-range interactions between molecules A and B from the venerable
expression~\cite{Casimir48}
\begin{equation}\label{eq:cp}
C_{6}(\text{A--B})=\frac{3}{\pi}\int_{0}^{\infty}
\alpha_{\text{A}}(iy)\alpha_{\text{B}}(iy)\,dy
\end{equation}
in which $i=\sqrt{-1}$.

\section{Results and discussion}\label{sec:res}

\subsection{How good are the additive models?}\label{sec:addres}

\begin{figure}
\includegraphics{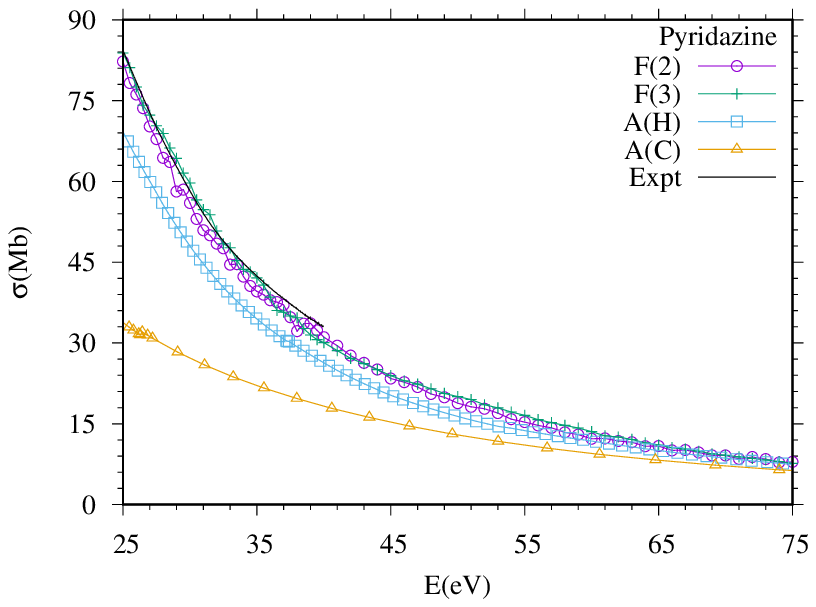}
\caption{\label{fig:Add2az}Photoabsorption cross-sections of pyridazine. The
experimental data is from Ref.~\onlinecite{Holland13}.}
\end{figure}

\begin{figure}
\includegraphics{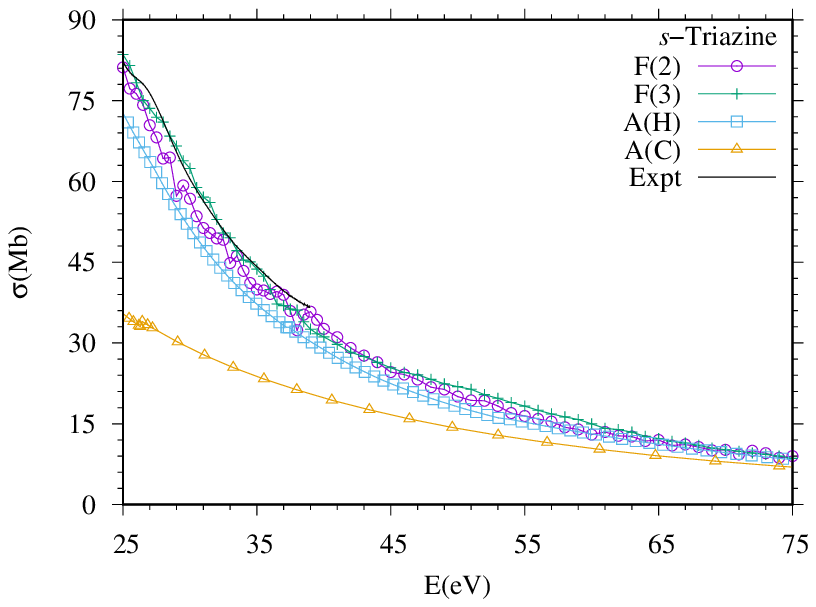}
\caption{\label{fig:Add3az}Photoabsorption cross-sections of $s$-triazine. The
experimental data is from Ref.~\onlinecite{Holland16}.}
\end{figure}

\begin{figure}
\includegraphics{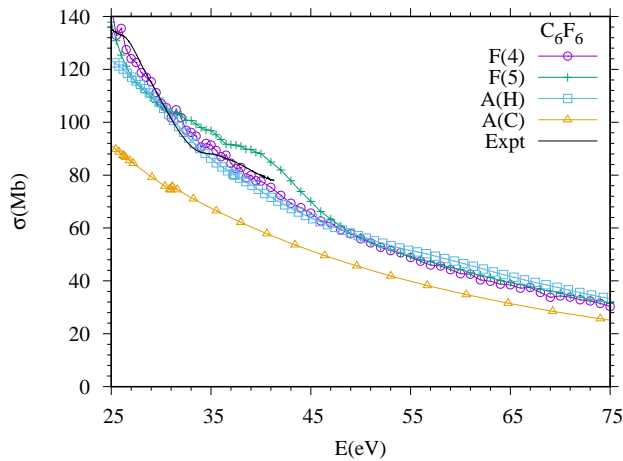}
\caption{\label{fig:Addhex}Photoabsorption cross-sections of hexafluorobenzene. The
experimental data is from Ref.~\onlinecite{Holland09}.}
\end{figure}

\begin{figure}
\includegraphics{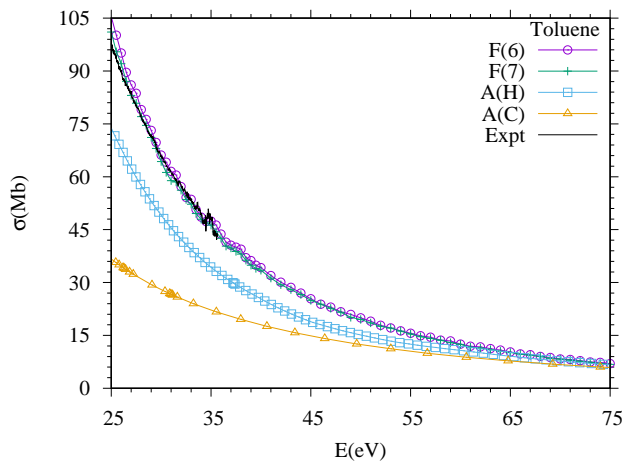}
\caption{\label{fig:Addtol}Photoabsorption cross-sections of toluene. The
experimental data is from Ref.~\onlinecite{Shaw98}.}
\end{figure}

\begin{figure}
\includegraphics{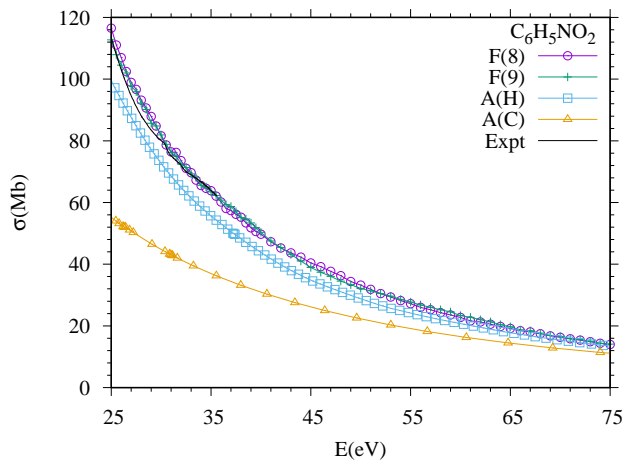}
\caption{\label{fig:Addnitro}Photoabsorption cross-sections of nitrobenzene. The
experimental data is from Ref.~\onlinecite{Cooper01}.}
\end{figure}

Two atom-additive models A(H) and A(C), two applicable fragment-additive models, and
experimentally measured cross-sections are compared with one another for
pyridazine,\cite{Holland13} $s$-triazine,\cite{Holland16}
hexafluorobenzene,\cite{Holland09} toluene,\cite{Shaw98} and
nitrobenzene\cite{Cooper01} in Figs.~\ref{fig:Add2az}, \ref{fig:Add3az},
\ref{fig:Addhex}, \ref{fig:Addtol}, and \ref{fig:Addnitro}, respectively.

The figures lead to five general observations.
\begin{enumerate}
\item The fragment-additive models F($m$) are seen to be reasonably accurate at the
lower energies where direct measurements are available. Hence the fragment models
F($m$) are very likely to be adequate for the somewhat higher energy range
$35\,\text{eV}\le{E}\le{200}\,\text{eV}$ in which they are used in this work.
\item Where direct measurements are available, the fragment-additive models F($m$)
are noticeably more accurate than both the atom-additive models A(H) and A(C) except
for hexafluorobenzene where A(H) is comparable to F(4) down to about 30$\,$eV\@.
\item As noted by Au \textit{et al.},\cite{Au94Mix} chemical intuition is not always
helpful in predicting which fragment model is more accurate. For example, F(3) is
more accurate than F(2) even though one might expect F(3) to be less accurate because
the triple bonds in the nitrogen and ethyne molecules are unrepresentative of the
bonding in the aromatic ring.
\item The A(H) model is consistently better than A(C) especially at lower energies,
perhaps because the free-atom data used in A(C) are based on the Hartree-Fock
approximation which neglects electronic correlation.
\item The A(H) model begins to converge to the fragment models at energies larger
than about 65$\,$eV\@. It can be used with confidence for $E>200\,\text{eV}$ where
the Brion data\cite{BrionDB} and hence our fragment models end.
\end{enumerate}

\subsection{DOSDs}\label{sec:dosd}

DOSDs for the eight (hetero)aromatic molecules were obtained with the methods and
data described in Sec.~\ref{sec:methodology}.  The number of distributions examined
ranged from 14 for pyridazine for which all the available experimental data comes
from a single source~\cite{Holland13} to 112 for benzene for which there is ample
data. The standard deviation $s$ of the scale factors measures the consistency of the
initial data and gives a rough idea of the accuracy of the DOSD\@. By this indicator,
the benzene DOSD is the most accurate; the value of $100s=0.5$ obtained for benzene
in this work is less than a quarter of the value $100s=2.2$ found\footnote{$100s$ is
the same as the quantity STD used in DOSD work by Meath and coworkers.} for the
benzene DOSD constructed decades earlier.\cite{Kumar92a} Interestingly, unlike the
older one, the best benzene DOSD of this work does not include any refractivity
constraints. The DOSD for pyrazine is the least accurate ($100s$=3.1) and the values
of $100s$ for the other six molecules are 1.2, 1.3, 1.5, 1.8, 1.8, and 2.2 for
toluene, nitrobenzene, $s$-triazine, pyridazine, hexafluorobenzene, and pyrimidine,
respectively. The integrated properties of the DOSDs are expected to be more accurate
than the point-wise distributions.\cite{Kumar08,Thakkar16DOSD}

\subsection{Polarizabilities and refractivities}\label{sec:pol}

The DOSD polarizabilities $\alpha(\omega)$ for smaller frequencies are expected to
have errors no larger than $\pm{1}\%$ for benzene, $\pm{3}\%$ for pyrazine, and
$\pm{2\%}$ for the remaining six molecules. Of the properties mentioned in
Sec.~\ref{sec:props}, the one studied the most is the static polarizability
$\alpha(0)$ usually denoted simply as $\alpha$. Consider benzene first. The current
DOSD value of $S(-2)=\alpha=68.19$ au is 0.6\% larger than and supersedes both the older
DOSD value\cite{Kumar92a} and identical recommended value\cite{Hohm13} of 67.79 au.
Table~\ref{tab:pol} compares the values of $\alpha$ obtained in this work with prior
theoretical computations. The density functional theory (DFT) result from the global
hybrid functional B3LYP is 1.6\% larger than the DOSD value whereas the $\alpha$
computed with the range-separated hybrid functional $\omega$B97X-D is within 0.1\% of
the DOSD value. The \emph{ab initio} value obtained\cite{Jorgensen20} with the CC3
coupled-cluster method is just 0.4\% too large for benzene.

\begin{table}
\caption{\label{tab:pol}Static electronic dipole polarizabilities $\alpha$ (in au).
Multiply by 0.1481847 to get the polarizability volume in {\AA}$^{3}$.}
\begin{ruledtabular}
\begin{tabular}{lcccc}
Molecule               & B3LYP\footnotemark[1] & $\omega$B97X-D\footnotemark[2]
& \textit{Ab initio}  & DOSD\footnotemark[6] \\
\hline
Benzene                & 69.31 & 68.26 & 68.49\footnotemark[3] & 68.19\footnotemark[7] \\
Pyridazine             & 58.93 & 58.00 & 58.73\footnotemark[3] & 57.09 \\
Pyrimidine             & 58.23 & 57.32 & 57.81\footnotemark[3] & 57.00 \\
Pyrazine               & 59.16 & 58.23 & 58.83\footnotemark[3] & 57.10 \\
$s$-Triazine           & 52.42 & 51.67 & 52.76\footnotemark[4] & 52.17 \\
Toluene                & 82.97 & 81.44 & 82.19\footnotemark[5] & 83.31 \\
C$_{6}$H$_{5}$NO$_{2}$ & 87.93 & 85.63 & 87.42\footnotemark[5] & 90.09 \\
C$_{6}$F$_{6}$         & 72.79 & 71.45 & ---                   & 71.40 \\
\end{tabular}
\end{ruledtabular}
\footnotetext[1]{Global hybrid density functional: B3LYP/aug-cc-pVTZ, Ref.~\onlinecite{Blair14pol}.}
\footnotetext[2]{Range-separated hybrid density functional: $\omega$B97X-D/aug-cc-pVTZ, Ref.~\onlinecite{CCCBDB20}.}
\footnotetext[3]{Coupled-cluster method: CC3/aug-cc-pVTZ, Ref.~\onlinecite{Jorgensen20}.}
\footnotetext[4]{Composite (hybrid) SDQ-MP4 value, Ref.~\onlinecite{Archibong94a}.}
\footnotetext[5]{2nd order M{\o}ller-Plesset perturbation theory: MP2/aug-cc-pVTZ, Ref.~\onlinecite{CCCBDB20}.}
\footnotetext[6]{This work.}
\footnotetext[7]{The old DOSD result of Kumar and Meath\cite{Kumar92a} is 67.79\,au.}
\end{table}

Now consider trends for the eight molecules. As
expected,\cite{Hickey14,Wu15DFT,Hait18} the B3LYP polarizabilities are consistently
larger than all the other values. Somewhat unexpectedly, the $\omega$B97X-D are
consistently closer to the DOSD values than the CC3 values are for benzene and the
diazines. The largest discrepancies, 3\% for CC3 and 2\% for $\omega$B97X-D, occur
for pyrazine whose DOSD is more poorly determined than the DOSDs for the other
molecules. Comparison of the DOSD polarizabilities with their free-atom additive
model counterparts, based on the exact\cite{Epstein26} polarizability for H and
accurate coupled-cluster values for the C, N, O, and F atoms,\cite{Das98pol2per}
shows that all eight molecules satisfy the minimum polarizability
principle\cite{Chattaraj96,Simon98} as more than 97\% of all molecules
do.\cite{Blair13MPP}

Next, turn to the frequency-dependent polarizability $\alpha(\omega)$ and to the
related molar refractivity $R(\omega)$ given by
\begin{equation}\label{eq:r}
R(\omega)=\frac{4{\pi}a_{0}^{3}}{3}N_{\text{A}}\alpha(\omega)
\end{equation}
in which $N_{\text{A}}$ is Avogadro's constant and $a_{0}$ is the Bohr radius.
Table~\ref{tab:refract} shows that the molar refractivities predicted by the benzene
DOSD are about 0.1\% larger than the measured values\cite{Ramaswamy1936} at all five
wave lengths. This close agreement comes about even though the best DOSD does not
have a refractivity constraint. The small discrepancies of 0.1\% imply that there is
a small incompatibility between the refractivity data\cite{Ramaswamy1936} and the low
energy photoabsorption cross-sections\cite{Rennie98,Etzkorn99,Feng02,Capalbo16} for
benzene. There are no measured gas-phase refractivities for the other seven
molecules.

\begin{table}
\caption{Molar refractivity $R$ (in cm$^{3}$mol$^{-1}$) as a function of wave length
$\lambda$ (in nm) for benzene. \label{tab:refract}}
\begin{ruledtabular}
\begin{tabular}{lcc}
$\lambda$ (nm) & DOSD\footnotemark[1] & Experiment\footnotemark[2] \\
\hline
 644.02   &  26.32           & 26.19       \\
 546.23   &  26.68           & 26.54       \\
 508.72   &  26.88           & 26.75       \\
 480.13   &  27.07           & 26.96       \\
 435.96   &  27.45           & 27.36       \\
\end{tabular}
\end{ruledtabular}
\footnotetext[1]{This work.}
\footnotetext[2]{Ref.~\onlinecite{Ramaswamy1936}.}
\end{table}

Table~\ref{tab:alfaomega} lists $\alpha(\omega)$ predicted by the DOSDs at two wave
lengths and compares them with CC3 values where available.\cite{Jorgensen20} Observe
that the CC3 values are consistently larger than the DOSD values at both wave
lengths. The CC3 and DOSD values differ by only 0.5\%, 3\%, 2\%, and 3\% for benzene,
pyridazine, pyrimidine, and pyrazine respectively. These discrepancies are within the
uncertainty of the DOSD polarizabilities except for pyridazine.

\begin{table}
\caption{Frequency-dependent polarizability $\alpha(\omega)$ (in au) as a
function of wave length $\lambda$ (in nm). \label{tab:alfaomega}}
\begin{ruledtabular}
\begin{tabular}{llcc}
Molecule   & Method               & $\lambda=632\,$nm & $\lambda=488\,$nm \\
\hline
Benzene                & CC3\footnotemark[1]  & 70.86 & 72.63 \\
                       & DOSD\footnotemark[2] & 70.51 & 72.26 \\
Pyridazine             & CC3\footnotemark[1]  & 60.62 & 62.07 \\
                       & DOSD\footnotemark[2] & 58.76 & 60.01 \\
Pyrimidine             & CC3\footnotemark[1]  & 59.76 & 61.11 \\
                       & DOSD\footnotemark[2] & 58.74 & 60.04 \\
Pyrazine               & CC3\footnotemark[1]  & 60.89 & 62.50 \\
                       & DOSD\footnotemark[2] & 59.09 & 60.64 \\
$s$-Triazine           & DOSD\footnotemark[2] & 53.65 & 54.77 \\
Toluene                & DOSD\footnotemark[2] & 86.16 & 88.31 \\
C$_{6}$H$_{5}$NO$_{2}$ & DOSD\footnotemark[2] & 93.65 & 96.46 \\
C$_{6}$F$_{6}$         & DOSD\footnotemark[2] & 73.32 & 74.77 \\
\end{tabular}
\end{ruledtabular}
\footnotetext[1]{Coupled-cluster method: CC3/aug-cc-pVTZ, Ref.~\onlinecite{Jorgensen20}.}
\footnotetext[2]{This work.}
\end{table}

\subsection{Sum rules and mean excitation energies}\label{sec:skik}

\begin{table*}
\caption{Dipole sum rules $S(k)$ (in au). $A(n)$ denotes $A{\times}10^{n}$.\label{tab:sk}}
\begin{ruledtabular}
\begin{tabular}{lccccccccc}
Molecule               & $S(2)$   & $S(1)$& $S(0)$& $S(-1)$ & $S(-2)$ & $S(-3)$ & $S(-4)$ & $S(-5)$ & $S(-6)$ \\
\hline
Benzene                & 1.947(4) & 297.4 &  42 & 39.93 & 68.19 & 152.8 & 420.6 & 1338  & 4659 \\
Pyridazine             & 2.515(4) & 338.9 &  42 & 36.03 & 57.09 & 118.9 & 304.8 & 912.6 & 3043 \\
Pyrimidine             & 2.517(4) & 339.6 &  42 & 35.86 & 57.00 & 120.4 & 315.4 & 968.0 & 3311 \\
Pyrazine               & 2.524(4) & 342.5 &  42 & 35.48 & 57.10 & 125.9 & 355.7 & 1211  & 4714 \\
$s$-Triazine           & 2.799(4) & 359.7 &  42 & 34.11 & 52.17 & 106.2 & 269.4 & 811.8 & 2801 \\
Toluene                & 2.280(4) & 350.8 &  50 & 48.37 & 83.31 & 186.9 & 515.0 & 1650  & 5828 \\
C$_{6}$H$_{5}$NO$_{2}$ & 4.624(4) & 553.9 &  64 & 54.18 & 90.09 & 209.3 & 629.1 & 2278  & 9339 \\
C$_{6}$F$_{6}$         & 1.215(5) & 1023  &  90 & 55.46 & 71.40 & 136.3 & 348.9 & 1083  & 3740 \\
\end{tabular}
\end{ruledtabular}
\end{table*}

The dipole sum rules $S(k)$ for $-6 \le k \le 2$ are listed in Table~\ref{tab:sk}.
Many important physical properties are related\cite{Thakkar16DOSD} to the $S(k)$. The
Taylor expansion of the frequency-dependent polarizability $\alpha(\omega)$ valid for
frequencies $\omega$ below the lowest excitation frequency $\omega_{1}$ is
\begin{equation}\label{eq:cauchy}
\alpha(\omega) = S(-2) + \omega^{2} S(-4) + \omega^{4} S(-6) + \cdots
\end{equation}
in which $S(-2)=\alpha(0)$ is the static electronic dipole polarizability and the
$S(-2k-2)$ with $k=1,2,\dots$ are called Cauchy moments. $S(2)$ is proportional to
the sum of the electron density values at the nuclei, and $S(-1)$ is related to the
total differential cross-section for inelastic scattering in collisions of
charged-particles with the molecule.

As can be verified in Table~\ref{tab:sk}, the DOSDs are constrained to satisfy the
KRT sum rule: $S(0)=N_{\text{e}}$ where $N_{\text{e}}$ is the number of electrons in
the molecule. The benzene $S(k)$ reported by Kumar and Meath\cite{Kumar92a} differ
from those in Table~\ref{tab:sk} by 4.6\% and 4.3\% for $k=2$ and 1, respectively,
and by $-1.8$\%, $-0.6$\%, 0.8\%, 2.3\%, 4.0\%, and 5.9\% for $k=-1$ to $k=-6$,
respectively.

The mean excitation energies $I(k)$ are listed in Table~\ref{tab:ik}. The average
energy associated with the total inelastic scattering cross-section for grazing
collisions of fast charged particles with the target species is $I(-1)$. The average
energies $I(0)$ and $I(1)$ which are respectively related to the average energy loss
(stopping power) and its mean fluctuation (straggling) in these collisions are
required in radiation damage theory. $I(2)$ is related to the Lamb shift.
\begin{table}
\caption{Mean excitation energies $I(k)$ (in eV). $A(n)$ denotes $A{\times}10^{n}$. \label{tab:ik}}
\begin{ruledtabular}
\begin{tabular}{lccccc}
Molecule              & $I(2)$  & $I(1)$& $I(0)$&$I(-1)$&$I(-2)$\\
\hline
Benzene               & 9.976(3) & 597.4 & 56.93 & 19.06 & 13.72 \\
Pyridazine            & 1.148(4) & 674.2 & 64.79 & 20.68 & 14.76 \\
Pyrimidine            & 1.146(4) & 673.5 & 65.17 & 20.72 & 14.63 \\
Pyrazine              & 1.140(4) & 672.0 & 66.32 & 20.73 & 14.24 \\
$s$-Triazine          & 1.205(4) & 708.9 & 69.17 & 21.60 & 15.18 \\
Toluene               & 9.908(3) & 594.4 & 55.95 & 18.80 & 13.67 \\
C$_{6}$H$_{5}$NO$_{2}$& 1.299(4) & 744.3 & 67.62 & 20.33 & 13.63 \\
C$_{6}$F$_{6}$        & 1.952(4) & 993.2 & 92.19 & 27.24 & 17.03 \\
\end{tabular}
\end{ruledtabular}
\end{table}
The benzene $I(k)$ reported by Kumar and Meath\cite{Kumar92a} differ from those in
Table~\ref{tab:ik} by 4.7\%, $-0.4$\%, 5.9\%, $-0.8$\% and $-1.2$\% for $k=2$, 1, 0,
$-1$, and $-2$ respectively.

\subsection{Dispersion coefficients \label{sec:disp}}

\begin{table}
\caption{Dispersion coefficients $C_{6}(\text{A--B})$ (in au). Multiplication of an
entry by 6934 yields a value in K\,{\AA}$^{6}$. \label{tab:disp}}
\begin{ruledtabular}
\begin{tabular}{lclc}
A--B   &   $C_{6}$ & A--B   &   $C_{6}$  \\
\hline
Benzene--Benzene             & 1765 & Pyrimidine--Toluene                & 1857 \\
Benzene--Pyridazine          & 1531 & Pyrimidine--C$_6$H$_5$NO$_2$       & 2009 \\
Benzene--Pyrimidine          & 1523 & Pyrimidine--C$_6$F$_6$             & 1775 \\
Benzene--Pyrazine            & 1507 & Pyrazine--Pyrazine                 & 1288 \\
Benzene--$s$-Triazine        & 1419 & Pyrazine--$s$-Triazine             & 1213 \\
Benzene--Toluene             & 2153 & Pyrazine--Toluene                  & 1838 \\
Benzene--C$_6$H$_5$NO$_2$    & 2328 & Pyrazine--C$_6$H$_5$NO$_2$         & 1988 \\
Benzene--C$_6$F$_6$          & 2050 & Pyrazine--C$_6$F$_6$               & 1756 \\
Pyridazine--Pyridazine       & 1330 & $s$-Triazine--$s$-Triazine         & 1143 \\
Pyridazine--Pyrimidine       & 1323 & $s$-Triazine--Toluene              & 1730 \\
Pyridazine--Pyrazine         & 1308 & $s$-Triazine--C$_6$H$_5$NO$_2$     & 1872 \\
Pyridazine--$s$-Triazine     & 1233 & $s$-Triazine--C$_6$F$_6$           & 1657 \\
Pyridazine--Toluene          & 1867 & Toluene--Toluene                   & 2625 \\
Pyridazine--C$_6$H$_5$NO$_2$ & 2020 & Toluene--C$_6$H$_5$NO$_2$          & 2838 \\
Pyridazine--C$_6$F$_6$       & 1785 & Toluene--C$_6$F$_6$                & 2498 \\
Pyrimidine--Pyrimidine       & 1315 & C$_6$H$_5$NO$_2$--C$_6$H$_5$NO$_2$ & 3070 \\
Pyrimidine--Pyrazine         & 1301 & C$_6$H$_5$NO$_2$--C$_6$F$_6$       & 2707 \\
Pyrimidine--$s$-Triazine     & 1226 & C$_6$F$_6$--C$_6$F$_6$             & 2416 \\
\end{tabular}
\end{ruledtabular}
\end{table}

The ten-term pseudospectra given in the supplementary material were used to
calculate the spherically averaged $C_{6}$ dispersion coefficients listed in
Table~\ref{tab:disp} for interactions between pairs of the eight (hetero)aromatic
molecules. The uncertainties in the $C_{6}$ coefficients are estimated to be in the
$\pm{4}$\% to $\pm{8}$\% range depending upon the quality of the underlying DOSDs for
the two species involved. The current value of 1765\,au for benzene is 2.4\% larger
than and supersedes the older value\cite{Kumar92a} of $C_{6}=1723\,\text{au}$.

Induced-dipole-induced-dipole $C_{6}$ coefficients are essential ingredients in the
construction of model, non-retarded, intermolecular potentials that are valid for all
intermolecular distances.\cite{Wheatley04} Moreover, they can assist in the
calibration of calculated intermolecular potentials.\cite{Szalewicz12} If the first
two dispersion coefficients $C_{6}$ and $C_{8}$ in the long-range interaction energy
expansion $V(R)=-C_{6}/R^{6}-C_{8}/R^{8}-C_{10}/R^{10}-\cdots$ are available, then
$C_{10}$ and higher-order coefficients can be calculated to good accuracy from simple
models.\cite{Thakkar74b,Thakkar88}

The supplementary material lists $C_{6}(\text{A--B})$ coefficients for unlike
(A$\ne$B) interactions in which A is one of the eight aromatic molecules considered
here and B is one of 75 other species for which published pseudospectra are
available.\footnote{The 75 species are comprised of 7 monatomic, 10 diatomic, 9
triatomic, 11 inorganic, and 38 organic species. The original references to their
published pseudospectra are enumerated in the supplementary material.} These 600
$C_{6}$ coefficients together with those in Table~\ref{tab:disp} constitute a
moderately large, self-consistent set of $C_{6}$ coefficients. This test set
facilitates a timely, contemporary assessment of a
well-regarded\cite{Tang69,Kramer70a,Zeiss77,Thakkar84c,Olney97} approximation:
\begin{equation}\label{eq:cr}
C_{6}(\text{A--B})=
\frac{2C_{6}(\text{A--A})C_{6}(\text{B--B})\alpha_{\text{A}}\alpha_{\text{B}}}
{C_{6}(\text{A--A})\alpha^{2}_{\text{B}}
+C_{6}(\text{B--B})\alpha^{2}_{\text{A}}} ,
\end{equation}
in which $\alpha_{\text{A}}$ and $\alpha_{\text{B}}$ are the mean static
polarizabilities of species A and B, respectively. Equation~(\ref{eq:cr}) is
sometimes called the Moelwyn-Hughes combination rule because it was published first
in his textbook.\cite{Moelwyn-Hughes57} However, it is quite likely that
Eq.~(\ref{eq:cr}) was known prior to that because it follows rather simply from the
London approximation\cite{London1930,*[{An English translation of
Ref.~\onlinecite{London1930} can be found in }] [{.}] Hettema00} for $C_{6}$. All
that is needed is to eliminate the Uns\"old average energies from Eq.~$(13')$ in
London's later paper\cite{London1937} using his Eq.~(13) from the same
paper.\cite{London1937}

The mean absolute percent deviation of Eq.~(\ref{eq:cr}) from the DOSD values is only
0.23\%. The predictions of Eq.~(\ref{eq:cr}) are in error by less than 1\% in 606
(96.4\%) of the 628 cases, and the error does not exceed 2.25\% in the worst case.
Interestingly, Eq.~(\ref{eq:cr}) is more likely to underestimate than overestimate
the DOSD value; it predicts an underestimate in 400 of the 628 cases examined.

\section{What next?}\label{sec:next}

The DOSDs constructed in this work for seven aromatic molecules plus the improved
DOSD for benzene increase significantly the number of molecules for which reliable
and complete DOSDs have been determined primarily from experimental photoabsorption
data. None of the new DOSDs incorporate a refractivity constraint unlike many, if not
most, DOSDs built in the past using various versions of the method used here. This is
encouraging because there are not too many more molecules for which gas-phase
refractivity data is available.\cite{Hohm13}

However, an examination of the MPI-Mainz UV/VIS Spectral Atlas\cite{KellerRudek13}
suggests that photoabsorption data from the absorption threshold to at least 30\,eV,
a range sufficient to construct a complete DOSD, is available for only about 20 more
molecules. Once DOSDs for them are constructed, as they surely will be in the near
future, what is the avenue for further progress? Obviously, experimental measurement
of photoabsorption cross-sections for more species is one.

A different path is to find robust, black-box-like, additive models that are
sufficiently accurate in the 10\,eV to 30\,eV range. That would open up the
possibility of constructing complete DOSDs for the hundreds of molecules for which
photoabsorption cross-sections are available\cite{KellerRudek13} in a restricted
energy range from the absorption threshold up to about 10\,eV\@. Free-atom and
fragment additive models are at levels 1 and 4 in the additive model
hierarchy.\cite{Thakkar12} Level 2 additive models using dressed atoms and level 3
models based on atoms that depend upon their environment deserve a closer look for
photoabsorption cross-sections. Preliminary work along these lines is under way.

\section*{Supplementary Material}\label{sec:supp}
See the supplementary material for tables of $C_{6}(\text{A--B})$ coefficients for
unlike (A$\ne$B) interactions in which A is one of the eight aromatic molecules
treated in this work and B is one of 75 other species, ten-term pseudospectra for the
eight aromatic molecules, and references to the published pseudospectra for the 75
other species.

\section*{Data availability}\label{sec:avail}
The data that supports the findings of this study are available within the article
and its supplementary material.

%

\end{document}